\documentclass[12pt]{iopart}

\usepackage{iopams}
\usepackage{dsfont}
\usepackage{graphics}

\newcommand{\avg}[1]{\langle{#1}\rangle}
\newcommand{\ket}[1]{\left|{#1}\right\rangle}
\newcommand{\bra}[1]{\left\langle{#1}\right|}

\begin{document}

\title{Toolbox for non-classical state calculations}
\author{Filippus S. Roux}
\ead{froux@nmisa.org}
\address{National Metrology Institute of South Africa, Meiring Naud{\'e} Road, Brummeria 0040, Pretoria, South Africa}
\begin{abstract}
Computational challenges associated with the use of Wigner functions to identify non-classical properties of states are addressed with the aid of generating functions. It allows the computation of the Wigner functions of photon-subtracted states for an arbitrary number of subtracted photons. Both the formal definition of photon-subtracted states in terms of ladder operators and the experimental implementation with heralded photon detections are analyzed. These techniques are demonstrated by considering photon subtraction from squeezed thermal states as well as squeezed Fock states. Generating functions are also used for the photon statistics of these states. These techniques reveal various aspects of the parameter dependences of these states.
\end{abstract}

\noindent{\it Keywords\/}: quantum optics, photon subtraction, Wigner function, generating function, squeezed thermal state

\section{Introduction}
\label{intro}

The advantage of quantum information systems over their classical equivalents can be related to the properties of the quantum states that serve as resources for such systems \cite{qadvant0,qadvant,qadvant1}. Therefore, the ``quantumness'' of these states is an important albeit elusive property, which can be defined or quantified in different ways.

Such quantum states are often distinguished from classical states based on some non-classical properties. An example of such a non-classical property is the presence of regions where the state's Wigner function in terms of the particle-number degrees of freedom\footnote{A Wigner function computed for the spatiotemporal degrees of freedom could have negative regions, but that would not indicate non-classical properties.} is negative. States with such negativity are useful resource states for quantum information processing \cite{resource,cvcount,gdistill}. The experimental techniques to prepare such non-classical states are reviewed in \cite{lvovsky}.

Most photonic states that are produced with existing technology have Gaussian Wigner functions, which do not contain negative regions on phase space. As a result, readily producible photonic states need additional processing to prepare non-classical photonic states. Negativity is generally obtained by applying some form of post-processing or post-selection on Gaussian states \cite{lvovsky}.

\begin{figure}[ht]
\centerline{\includegraphics{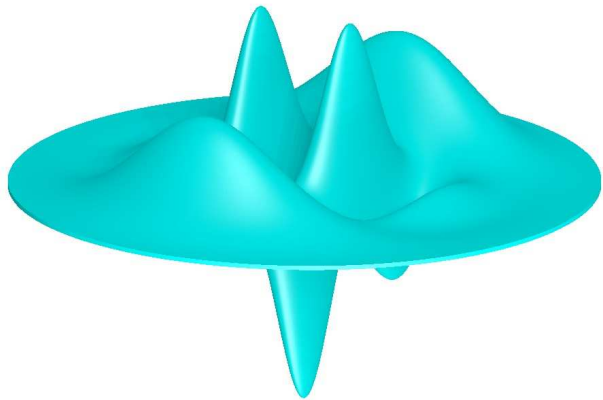}}
\caption{Wigner function of a squeezed vacuum states with five photons subtracted (not to scale).}
\label{catsub}
\end{figure}

One such process is {\em photon subtraction}, where one or more photons are removed from a given input state \cite{biswas}. Such photon-subtracted states can have regions on phase space where the Wigner function is negative. However, the input state needs to be a special state to produce such negativity \cite{trepstheorem}. Photon-subtracted squeezed vacuum states have Wigner functions with negative regions. Their Wigner functions resemble those of Schr{\"o}dinger cat states. For example, the three-dimensional view of the Wigner function of a squeezed vacuum state with five photons subtracted, shown in figure~\ref{catsub}, demonstrates its resemblence with the Wigner function of a Schr{\"o}dinger cat state. For a Schr{\"o}dinger cat state produced as the superposition of two coherent states, its Wigner function consists of two Gaussian functions displaced to geometrically opposite locations with respect to the origin and a Gaussian function at the origin modulated by oscillations along a direction perpendicular to the displacements. The Wigner function in figure~\ref{catsub} has the same qualitative features. Photon subtraction from squeezed vacuum states has indeed been proposed as a way to produce Schr{\"o}dinger cat states \cite{dakna}.

Another way to produce Wigner negativity, closely related to photon subtraction, is the opposite process: {\em photon addition} \cite{zavatta}. The formal definition implies an interchange of creation and annihilation operators, and the experimental implementation replaces the beam splitter with an optical parametric amplifier, which can be seen as replacing the SU(2) process by a SU(1,1) process. Here, we focus on photon subtraction since it suffices to demonstrate the use of generating functions. However, similar techniques can be used to aid the analytical calculations associated with photon addition.

While a large set of photon-subtracted states exists, depending on the number of photons that are subtracted and the type of input state that is being used, many of these cases are challenging to analyze, especially when multiple photons are subtracted or when the input state is a mixed state. Such challenges are often addressed with the aid of numerical simulations \cite{bright,trepstheorem}. To address these challenges without resorting to numerical simulations, we propose a generating function approach for such analyses.

Formally, photon subtraction involves the application of ladder operators to the state $\hat{\rho}\rightarrow \hat{a} \hat{\rho} \hat{a}^{\dag}$. In practice, the photon subtraction process is performed by detecting a specific number of photons in a portion of the state, which heralds the existence of the photon-subtracted state in the remainder \cite{lvovsky}.

Here, we investigate the formal process, as well as its experimental implementation. We compute the Wigner functions after the subtraction process is applied to an input state to see if it contains regions where the Wigner function is negative. While many of these states are qualitatively similar, their quantitative properties are different. For coherent states, neither the formal subtraction process, nor any experimental implementation thereof can produce a Wigner function with negative regions. Therefore, the input state needs to be squeezed \cite{trepstheorem}.

The formal photon-subtracted squeezed vacuum state has been compared to a squeezed Fock state \cite{lvovsky}. Here, we use a generating function for the Wigner functions of Fock states to investigation this comparison.

We consider only the particle-number degrees of freedom in these analyses. However, the approach presented here can also be applied in cases where the spatiotemporal degrees of freedom are included. Thus, one can generalize it to continuous variables in terms of the symplectic formalism \cite{contvar1,contvar2} for a finite number of discrete modes. Alternatively, one can use it in a Wigner functional formalism \cite{ipfe,entpdc,queez}, which includes all the spatiotemporal degrees of freedom without any restrictions or truncations.

\section{Generating functions}
\label{genfunk}

The operations that are performed to produce photon-subtracted states can either be modeled formally in terms of ladder operators or practically in terms of projection operators. Both these approaches are generalized to represent the subtraction of an arbitrary number of photons. Here, we alleviate the burden of such calculations for arbitrary numbers of photon subtractions by introducing generating functions.

To determine whether the Wigner functions of photon-subtracted states have negative regions on phase space, the calculations can be converted into the Moyal formalism \cite{groenewold,moyal,psqm} where all the operators are represented in terms of their Wigner functions. Products of operators then become star products of their Wigner functions.

Instead of using the Wigner functions of the ladder operators and projection operators in these phase space calculations, we use the Wigner functions of the generating functions of these operators. It alleviates the phase space calculations by producing generating functions for the Wigner functions of states with arbitrary numbers of subtracted photons.

\subsection{The concept of a generating function}

A generating function is a form of mathematical tagging, where multiple mathematical entities are combined into one mathematical entity in which the different entities are individually tagged in some way. The tagging makes it possible to extract the individual entities from the combined entity. In the case of a generating function, the tagging is done with the aid of a generating parameter, raised to different powers and then multiplied with the different entities (functions) before being summed. For example,
\begin{equation}
\mathcal{G}(J) = \sum_{n=0}^{\infty} J^n f_n ,
\end{equation}
where $J$ is the generating parameter and $f_n$ represents the different entities or functions in this case. The individual entities in a generating function are extracted with the aid of derivatives in the same way that one would calculate the terms in a Taylor series expansion. For the above example,
\begin{equation}
f_n = \frac{1}{n!} \left. \partial_J^n \mathcal{G} \right|_{J=0} .
\end{equation}
The benefit of such combined entities is that one can apply a linear operation to all the entities by applying the linear operation once to the combined entity. For instance, applying a linear operation represented by $L\{\cdot\}$ to the above example, we get
\begin{equation}
L\{\mathcal{G}(J)\} = \sum_{n=0}^{\infty} J^n L\{f_n\} .
\end{equation}
The results of the linear operation on individual entities can be extracted from this result in the same way as the individual entities are extracted from the generating function.

\subsection{Generating functions for ladder operators}

The generating functions for multiple ladder operators are given by
\begin{equation}
\hat{G} = \exp(\eta^* \hat{a}) ~~~ {\rm and} ~~~
\hat{G}^{\dag} = \exp(\hat{a}^{\dag}\eta) ,
\label{geedef}
\end{equation}
where $\hat{a}$ and $\hat{a}^{\dag}$ are ladder operators (annihilation and creation operators, respectively), and $\eta=J+\rmi K$ is a complex generating parameter. The derivatives with respect to these parameters can be defined in terms of derivatives with respect to their real and imaginary parts:
\begin{equation}
\partial_{\eta} \equiv \frac{1}{2}\left(\frac{\partial}{\partial J} - \rmi \frac{\partial}{\partial K}\right) ~~~ {\rm and} ~~~
\partial_{\eta^*} \equiv \frac{1}{2}\left(\frac{\partial}{\partial J} + \rmi \frac{\partial}{\partial K}\right) .
\end{equation}
It then follows that
\begin{equation}
\partial_{\eta} \eta^* = \partial_{\eta^*} \eta = 0 .
\end{equation}
As a result, the complex generating parameter and its complex conjugate act as independent generating parameters. Multiple ladder operators are produced from these generating functions by
\begin{equation}
\left(\hat{a}\right)^n = \left. \partial_{\eta^*}^n \hat{G} \right|_{\eta^*=0} ~~~ {\rm or} ~~~
\left(\hat{a}^{\dag}\right)^n = \left. \partial_{\eta}^n \hat{G}^{\dag} \right|_{\eta=0} .
\label{difflad}
\end{equation}

The Wigner functions of the generating functions in equation~(\ref{geedef}) are given by
\begin{equation}
W_{\hat{G}} = \exp(\eta^*\alpha) ~~~ {\rm and} ~~~
W_{\hat{G}^{\dag}} = \exp(\alpha^*\eta) ,
\label{genladwig}
\end{equation}
where
\begin{equation}
\alpha \equiv \frac{1}{\sqrt{2}} (q+\rmi p) .
\label{anaqp}
\end{equation}
The Wigner functions for specific powers of the ladder operators are obtained by applying equation~(\ref{difflad}) to the Wigner functions in equation~(\ref{genladwig}). For example, by replacing $\hat{G}\rightarrow W_{\hat{G}}$ in equation~(\ref{difflad}), we obtain $\alpha^n$.

\subsection{Generating function for projection operators}
\label{projgen}

The photon-number-resolving detection process is represented by projection operators for fixed numbers of photons
\begin{equation}
\hat{P}_n = \ket{n}\bra{n} .
\end{equation}
The probability to detect $n$ photons is given by
\begin{equation}
P(n) = \tr\{\hat{P}_n\hat{\rho} \} = \bra{n}\hat{\rho}\ket{n} .
\end{equation}
However, it is more convenient to perform the calculations with the aid of a generating function for these projection operators. It is defined by
\begin{equation}
\mathcal{P} = \sum_{n=0}^{\infty} J^n \hat{P}_n = \sum_{n=0}^{\infty} \ket{n} J^n \bra{n} ,
\end{equation}
where $J$ is the generating parameter, so that
\begin{equation}
\hat{P}_n = \left. \frac{1}{n!} \partial_J^n \mathcal{P} \right|_{J=0} .
\end{equation}
When the generating parameter is set equal to 1, the generating function reproduces the identity
\begin{equation}
\left. \mathcal{P} \right|_{J=1} = \sum_{n=0}^{\infty} \ket{n} \bra{n} = \mathds{1} ,
\end{equation}
where $\mathds{1}$ is the identity operator.

A generating function for the probability to detect $n$ photons from a state $\hat{\rho}$ is obtained from the trace
\begin{equation}
\mathcal{F} = \tr\{\mathcal{P}\hat{\rho}\} = \sum_{n=0}^{\infty} J^n \tr\{\hat{P}_n\hat{\rho}\} = \sum_{n=0}^{\infty} J^n P(n) .
\end{equation}
This generating function does not only provide the means to produce the individual probabilities. It can also be used to compute the expectation value for the number of photons in the state. To calculate this expectation value, we perform
\begin{equation}
\left. \partial_J \mathcal{F} \right|_{J=1} = \sum_{n=0}^{\infty} n P(n) = \langle n\rangle .
\end{equation}
The variance in the number of photons in the state, which is given by
\begin{equation}
\sigma^2 = \langle n^2\rangle - \langle n\rangle^2 ,
\end{equation}
can be calculated with the aid of the second moment, which is directly obtained from the generating function as
\begin{equation}
\left. \partial_J \left(J \partial_J \mathcal{F}\right) \right|_{J=1} = \sum_{n=0}^{\infty} n^2 P(n) = \langle n^2\rangle .
\end{equation}

Since we intend to perform these calculations with the aid of Wigner functions, we need the Wigner function of the generating function for these projection operators. It is given by
\begin{equation}
\mathcal{W}_{\mathcal{P}} = \frac{2}{1+J} \exp\left(-2|\alpha|^2 \frac{1-J}{1+J} \right) .
\label{wigproj}
\end{equation}
Since the projection operator is also the density operator for a Fock state, the Wigner function in equation~(\ref{wigproj}) also serves as a generating function for the Wigner functions of Fock states. This Wigner function shares some properties with the original generating function. By setting $J=1$, we obtain the sum over the Wigner functions of all the projection operators. It gives
\begin{equation}
\mathcal{W}_{\mathcal{P}}(J=1) = 1,
\end{equation}
which is the Wigner function for the identity. When we apply a derivative to the Wigner function in equation~(\ref{wigproj}) before setting $J=1$, the result is the Wigner function for the number operator
\begin{equation}
\left. \partial_J \mathcal{W}_{\mathcal{P}} \right|_{J=1} = |\alpha|^2-\case{1}{2} \equiv W_{\hat{n}} ,
\end{equation}
which produces the expectation value for the number of photons in a state when it is traced with the Wigner function of that state.


\section{Mixed squeezed vacuum states}
\label{bogo}

As an example, we'll consider the case where the input state is a mixed squeezed vacuum state. Such a state is obtained when a homogeneous loss (as found when passing the state through a neutral density filter) is applied to a pure squeezed vacuum state. It can also be produced as a squeezed thermal state.

The Wigner function for a pure squeezed vacuum state is
\begin{equation}
W_{{\rm sq}} = 2\exp\left(-2|\alpha|^2 A-\alpha^2 B^*-\alpha^{*2} B\right) ,
\label{squ}
\end{equation}
where $A^2-|B|^2=1$. The constants are parameterized in terms of a complex-valued squeezing parameter $\xi=|\xi|\exp(\rmi\varphi)$, by
\begin{equation}
A = \cosh(2|\xi|) , ~~~~~
B = \exp(\rmi\varphi)\sinh(2|\xi|) .
\label{defab}
\end{equation}
They are related to the average number of photons in the squeezed vacuum state $\avg{n}$ by
\begin{equation}
A = 1+2\avg{n} , ~~~~~
B = 2\exp(\rmi\varphi)\sqrt{\avg{n}}\sqrt{\avg{n}+1} .
\label{defabn}
\end{equation}
Henceforth, we'll assume that $\varphi=0$, which means the Wigner function of the squeezed state is elongated along the $p$-direction and squeezed along the $q$-direction.

When a pure squeezed vacuum state passes through a neutral density filter with transmission coefficient $t<1$, the Wigner function becomes
\begin{equation}
W_{{\rm msq}} = \frac{2}{\sqrt{1+2ta(A-1)}}\exp\left[\frac{-2|\alpha|^2 (a+tA)-t\alpha^2 B^*-t\alpha^{*2} B}{1+2ta(A-1)}\right] ,
\label{msqu}
\end{equation}
where $a=1-t$ is the absorption coefficient. The resulting state is mixed, having a purity of
\begin{equation}
{\rm purity} = \frac{1}{\sqrt{1+2ta(A-1)}} .
\end{equation}

Another way to produce a mixed squeezed vacuum state is to apply squeezing to a thermal state. The Wigner function for a thermal state is
\begin{equation}
W_{{\rm th}} = 2T\exp\left(-2T|\alpha|^2\right) ,
\end{equation}
where $T<1$ is the purity of the state. For $T=1$, we obtain the Wigner function of a vacuum state.

The squeezing process is done with the aid of a Bogoliubov transformation. It is a unitary process that transforms the arguments of the Wigner function of the state on which it is applied. Hence, the Bogoliubov transformation produces
\begin{equation}
W_{\hat{\rho}}(\alpha) \rightarrow W_{\hat{\rho}}(U\alpha+V\alpha^*) ,
\end{equation}
where $U$ and $V$ are constants expressed in terms of the complex-valued squeezing parameter $\xi=|\xi|\exp(\rmi\varphi)$ by
\begin{equation}
U = \cosh(|\xi|) , ~~~~~
V = \exp(\rmi\varphi)\sinh(|\xi|) ,
\end{equation}
so that $U^2-|V|^2=1$.

After applying a Bogoliubov transformation to the thermal state, we obtain a squeezed thermal state with a Wigner function expressed as
\begin{equation}
W_{{\rm sts}} = 2T\exp\left(-2T|\alpha|^2 A_0-T\alpha^2 B_0^*-T\alpha^{*2} B_0\right) .
\label{sqth}
\end{equation}
The constants $A_0$, $B_0$, and $B_0^*$ are produced by the Bogoliubov transformation and are given in equation~(\ref{defab}) in terms of the squeezing parameter.

The expressions in equations~(\ref{msqu}) and (\ref{sqth}) are formally equivalent. It can be shown by setting
\begin{equation}
\eqalign{
T = \frac{1}{\sqrt{1+2ta(A-1)}} , \cr
A_0 = \frac{a+tA}{\sqrt{1+2ta(A-1)}} , \cr
B_0 = \frac{tB}{\sqrt{1+2ta(A-1)}} , \cr
B_0^* = \frac{tB^*}{\sqrt{1+2ta(A-1)}} , }
\end{equation}
in equation~(\ref{sqth}) to obtain equation~(\ref{msqu}). Since the Bogoliubov transformation is a unitary process, the purity of the state is not affected by the transformation.


\section{Formal photon subtraction}
\label{formal}

The formal representation of the process for the subtraction of $n$ photons from a state $\hat{\rho}$ is defined by
\begin{equation}
\hat{\rho}_{{\rm fs}} = \frac{\left(\hat{a}\right)^n \hat{\rho} \left(\hat{a}^{\dag}\right)^n}
{\tr\left\{\left(\hat{a}\right)^n \hat{\rho} \left(\hat{a}^{\dag}\right)^n\right\}} .
\end{equation}
The denominator serves to normalize the state after the subtraction process, which does not preserve the normalization of the state.

The Wigner function of the state obtained from the formal photon subtraction process is obtained with the aid of a triple star product. It involves the Wigner functions of the input state and those of the generating functions for the ladder operators given in equation~(\ref{genladwig}). By computing this triple star product, we obtain a generating function for the unnormalized Wigner function of the photon-subtracted states in terms of the Wigner function of the input state. The expression reads
\begin{eqnarray}
\mathcal{W}_{\hat{G}\hat{\rho}\hat{G}^{\dag}} & = \int \exp[(\alpha^*-\alpha_1^*)\alpha_2-\alpha_2^*(\alpha-\alpha_1)] W_{\hat{G}}[\case{1}{2}(\alpha_1+\alpha+\alpha_2)] \nonumber \\
& ~~~ \times W_{\hat{\rho}}(\alpha_1)
W_{\hat{G}^{\dag}}[\case{1}{2}(\alpha_1+\alpha-\alpha_2)]\ \rmd^2\alpha_1\ \rmd^2\alpha_2 \nonumber \\
 & = \exp\left(\eta^*\alpha+\alpha^*\eta+\case{1}{2} |\eta|^2\right)
 W_{\hat{\rho}}\left(\alpha+\case{1}{2} \eta\right) ,
\label{ps}
\end{eqnarray}
where
\begin{equation}
\rmd^2\alpha_n = \frac{1}{2\pi}\ \rmd q_n\ \rmd p_n .
\end{equation}

For single-photon subtraction, we compute
\begin{equation}
\left. \partial_{\eta} \partial_{\eta^*} \mathcal{W}_{\hat{G}\hat{\rho}\hat{G}^{\dag}} \right|_{\eta^*=\eta=0}
 = \case{1}{2}\left( \case{1}{2}\partial_{\alpha}\partial_{\alpha^*} + \alpha\partial_{\alpha}
 + \alpha^*\partial_{\alpha^*}+1+2|\alpha|^2\right) W_{\hat{\rho}} ,
\label{formeel}
\end{equation}
which gives a differential operation that produces the (unnormalized) Wigner function for the single-photon-subtracted state when it is applied to the Wigner function of an arbitrary input state. In a similar way, one can compute the differential operations required for larger numbers of photon subtractions.

The trace of the generating function
\begin{equation}
\mathcal{W}_{\tr\{\hat{G}\hat{\rho}\hat{G}^{\dag}\}} = \int \mathcal{W}_{\hat{G}\hat{\rho}\hat{G}^{\dag}}\ \rmd^2\alpha ,
\label{trfs}
\end{equation}
gives a generating function for the traces of the photon-subtracted states. These traces serve as the inverses of the required normalization constants. The Wigner functions for the normalized $n$-photon-subtracted states (according to the formal process) are then given by
\begin{equation}
W_{\hat{\rho}-n} = \left. \frac{\partial_{\eta^*}^n\partial_{\eta}^n\mathcal{W}_{\hat{G}\hat{\rho}\hat{G}^{\dag}}}
{\partial_{\eta^*}^n\partial_{\eta}^nW_{\tr\{\hat{G}\hat{\rho}\hat{G}^{\dag}\}}} \right|_{\eta=\eta^*=0} .
\label{fstate}
\end{equation}

\subsection{Formal photon-subtracted squeezed thermal states}

Applying equation~(\ref{ps}) to the Wigner function in equation~(\ref{sqth}), we obtain a generating function for photon subtraction from a squeezed thermal state, according to the formal definition of the process. It is given by
\begin{eqnarray}
\mathcal{W}_{{\rm fs-sts}} = & 2T\exp\left[-2T|\alpha+\case{1}{2}\eta|^2 A
-T(\alpha+\case{1}{2}\eta)^2 B^*-T(\alpha^*+\case{1}{2}\eta^*)^2 B \right. \nonumber \\
& \left. +\eta^*\alpha+\alpha^*\eta+\case{1}{2} |\eta|^2\right] ,
\label{gensqth}
\end{eqnarray}
where we dropped the subscript 0 of the constants. The generating function for the traces of these states, obtained by computing the trace of the generating function in equation~(\ref{gensqth}), is given by
\begin{equation}
\mathcal{W}_{{\rm tr\{fs-sts\}}} = \exp\left(\frac{2|\eta|^2 (A-T)-\eta^2 B^*-\eta^{*2} B}{4T}\right) .
\label{trsqth}
\end{equation}
A state with $n$ subtracted photons is obtained from this generating function with the aid of equation~(\ref{fstate}). For one subtracted photon, the normalized Wigner function is
\begin{eqnarray}
W_{{\rm sts}-1} = & \left[\left(1-T^2\right)|\alpha|^2
+(TA-1)\left(2T|\alpha|^2A+T\alpha^2 B^*+T\alpha^{*2} B-\case{1}{2}\right)\right] \nonumber \\
& \times \frac{4T^2}{A-T} \exp\left(-2T|\alpha|^2A-T\alpha^2 B^*-T\alpha^{*2} B\right) .
\label{eenfs}
\end{eqnarray}
The region at the center, where $\alpha=0$ is negative, provided that $TA>1$.

\begin{figure}[ht]
\centerline{\includegraphics{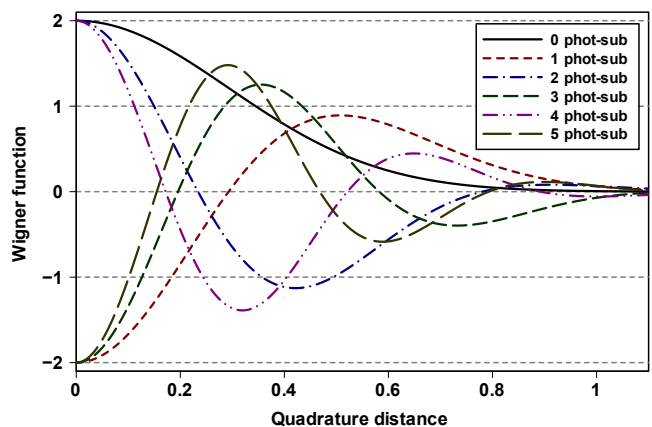}}
\caption{Wigner functions of formal photon-subtracted squeezed vacuum states ($T=1$) with zero to five subtracted photons, plotted along the squeezed direction. The input squeezed vacuum state has $\avg{n}=1$, so that $A=3$ and $B=B^*=2.83$.}
\label{for1p0}
\end{figure}

\begin{figure}[ht]
\centerline{\includegraphics{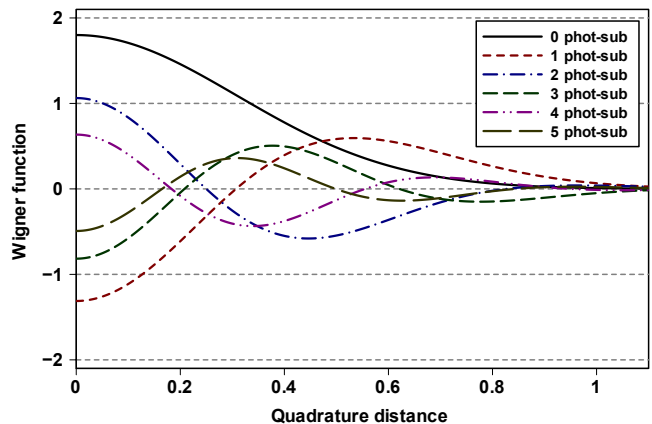}}
\caption{Wigner functions of formal photon-subtracted squeezed thermal states with zero to five subtracted photons, plotted along the squeezed direction for $T=0.9$, $A=3$ and $B=B^*=2.83$.}
\label{for0p9}
\end{figure}

The Wigner functions of formal photon-subtracted squeezed vacuum states ($T=1$) and photon-subtracted squeezed thermal states with purity $T=0.9$, for zero to five subtracted photons, are shown in figures~\ref{for1p0} and \ref{for0p9}, respectively. In both figures, the parameter values for $A$, $B$, and $B^*$ are the same, given by equation~(\ref{defabn}) for a squeezed vacuum state with $\avg{n}=1$. All the curves are plotted along the line where $p=0$ as a function of $q$, which is the squeezed direction. The curves in figures~\ref{for1p0} and \ref{for0p9} show the positive and negative regions produced by the photon subtractions, each having a width that is smaller than the width of a coherent state. The number of negative regions is equal to the number of subtracted photons. For an odd number of subtracted photons, there is always a negative region located at the origin. In figure~\ref{for1p0}, the amplitudes of the curves start from the same maximum or minimum value, but in figure~\ref{for0p9} the amplitudes decay for larger numbers of subtracted photons. This trend indicates that mixing has a detrimental effect on the amplitudes of the oscillations produced by the photon subtractions. The extent of the sensitivity of these oscillations to a reduction in the purity, is made apparent by the fact that, even for the slightly reduced value of $T=0.9$, the decay in the amplitudes of the oscillations is already significant.

\section{Heralded photon subtraction}
\label{herald}

\subsection{Experimental process}
\label{fotsubexp}

The formal definition of the photon subtraction process does not correspond directly to a physical process; the physical implementation of the photon subtraction process is different from its formal description. It uses projective measurements in conjunction with a beam splitter to simulate the action of the ladder operators. Nevertheless, one can model the physically implemented photon subtraction process to investigate it analytically and compare it with the formal description.

The experimental implementation of a photon subtraction process involves a heralded process whereby the detection of a specific number of photons in a portion of the state signals the existence of the required photon-subtracted state in the remainder. For this purpose, a portion of the state is separated from the remainder of the state with the aid of an unbalanced beam splitter. The portion that is separated off, is then subjected to a measurement to detect a specific number of photons. Ideally, a photon-number-resolving detector is needed for this purpose, but the generating function approach allows one to investigate the case where the detector is not photon-number resolving.

The two input ports of the beam splitter respectively receive the input state and a vacuum state. The unitary process that represents the beam splitter transforms the arguments of the product of the Wigner functions of these two states, leading to
\begin{equation}
W_{\hat{\rho}}(\alpha)W_{{\rm vac}}(\beta) \rightarrow
W_{\hat{\rho}}(\sqrt{1-\zeta}\alpha+\rmi \sqrt{\zeta}\beta) W_{{\rm vac}}(\sqrt{1-\zeta}\beta+\rmi \sqrt{\zeta}\alpha) ,
\label{wigbstra}
\end{equation}
where $\zeta$ represents the reflectivity of the beam splitter. The measurement is done on the portion of the state represented by $\beta$.

The projective measurement is performed with the aid of the generating function for the projection operators in equation~(\ref{wigproj}), with $\alpha\rightarrow\beta$. After multiplying it with the Wigner function after the beam splitter, we compute the trace by integrating over $\beta$:
\begin{eqnarray}
\mathcal{W}_{{\rm hs}} = & \int W_{\hat{\rho}}(\sqrt{1-\zeta}\alpha+\rmi \sqrt{\zeta}\beta)
W_{{\rm vac}}(\sqrt{1-\zeta}\beta+\rmi \sqrt{\zeta}\alpha) \nonumber \\
& \times \frac{2}{1+J} \exp\left(-2|\beta|^2 \frac{1-J}{1+J} \right)\ \rmd \beta .
\label{meetgen}
\end{eqnarray}
The result is a generating function for the Wigner functions of the states that are heralded by the detection of certain numbers of photons. However, these Wigner functions are not normalized, because the heralding process does not preserve the trace. The trace of the generating function in equation~(\ref{meetgen}), which implies an integration over $\alpha$, provides a generating function for the traces of states,
\begin{equation}
\mathcal{W}_{\tr\{{\rm hs}\}} = \int \mathcal{W}_{{\rm hs}}\ \rmd^2\alpha ,
\label{trhs}
\end{equation}
as it was done for the formal process in equation~(\ref{trfs}). The individual normalized Wigner functions are then generated by
\begin{equation}
W_{\hat{\rho}-n} = \left. \frac{\partial_J^n \mathcal{W}_{{\rm hs}}}{\partial_J^n \mathcal{W}_{\tr\{{\rm hs}\}}} \right|_{J=0} .
\end{equation}
like the way it is done for the formal process in equation~(\ref{fstate}).

For the evaluation of the integral in equation~(\ref{meetgen}), we need to specify the Wigner function of the input state $W_{\hat{\rho}}$. However, it is often assumed that the beam splitter separates off only a small portion of the state, so that $\zeta\ll 1$. In that case, we can expand equation~(\ref{meetgen}) up to first order in $\zeta$. The resulting expression then allows us to evaluate the integral over $\beta$, without having to specify the input state. The result reads
\begin{eqnarray}
\mathcal{W}_{{\rm hs}} \approx & \case{1}{2} \zeta
\left[\case{1}{2} (1+J) \partial_{\alpha}\partial_{\alpha^*}
+ J\alpha\partial_{\alpha}
+ J\alpha^*\partial_{\alpha^*}\right] W_{\hat{\rho}} \nonumber \\
& + \left[1+\case{1}{2} (1+J)\zeta-(1-J) \zeta |\alpha|^2\right] W_{\hat{\rho}} ,
\end{eqnarray}
It only allows single-photon detection at this level of the expansion, which leads to
\begin{equation}
\left. \partial_J \mathcal{W}_{{\rm hs}} \right|_{J=0} \approx \case{1}{2} \zeta \left(\case{1}{2} \partial_{\alpha}\partial_{\alpha^*}
+ \alpha\partial_{\alpha} + \alpha^*\partial_{\alpha^*} + 1 + 2|\alpha|^2\right) W_{\hat{\rho}} .
\label{kleins}
\end{equation}
A comparison with the differential operation in equation~(\ref{formeel}) produced by the formal process shows that, in the limit $\zeta\ll 1$, the two processes produce the same state for single photon subtraction.

The detection of $n$ photons assumes that the detector is photon-number resolving. However, it is also possible to use a non-photon-number-resolving detector. The ensemble averaged state that would be produced by heralding the detection of an arbitrary number of photons is obtained by removing the probability to observe the vacuum. In terms of the generating function, it is given by
\begin{equation}
W_{{\rm npnr}} = \mathcal{N} \left(\left.\mathcal{W}_{{\rm hs}}\right|_{J=1} - \left.\mathcal{W}_{{\rm hs}}\right|_{J=0}\right) ,
\end{equation}
where $\mathcal{N}$ is a required normalization factor. Under the assumption that $\zeta\ll 1$, the state that is obtained from a non-photon-number-resolving detection is the same as the one obtained in equation~(\ref{kleins}).

\subsection{Heralded photon-subtracted squeezed thermal states}

The generating function for heralded photon subtraction from a squeezed thermal state is obtained by substituting the Wigner function for the squeezed thermal state, given in equation~(\ref{sqth}), into equation~(\ref{meetgen}) and evaluating the integral over $\beta$. The resulting generating function reads
\begin{eqnarray}
\fl \mathcal{W}_{{\rm hs-sts}} = & \frac{2T}{\sqrt{\mathcal{K}}}
\exp\left[-\frac{(1-\zeta)T}{\mathcal{K}}\left(2|\alpha|^2A+\alpha^2 B^*+\alpha^{*2} B\right)
 -\frac{\zeta}{\mathcal{K}}(1-J+T^2+T^2J)|\alpha|^2 \right. \nonumber \\
\fl & \left. - \frac{\left(1-J^2\right)\zeta^2}{2\mathcal{K}}|\alpha|^2(2TA-1-T^2) \right] .
\label{hs}
\end{eqnarray}
where
\begin{equation}
\mathcal{K} \equiv 1+(1+J)(TA-1)\zeta-\case{1}{4}(1+J)^2(2TA-1-T^2)\zeta^2 .
\end{equation}
The generating function for the associated traces reads
\begin{equation}
\mathcal{W}_{{\rm tr\{hs-sts\}}} = \left[1+\frac{(1-J)(A-T)\zeta}{T}-\frac{(1-J)^2(2TA-1-T^2)\zeta^2}{4T^2}\right]^{-1/2} .
\end{equation}

The expressions of the normalized Wigner functions of the photon-subtracted states are more complicated for the heralded process than for the formal process. In the limit of a small reflectivity $\zeta\rightarrow 0$, the expression for one subtracted photon becomes the same as in equation~(\ref{eenfs}). For larger numbers of subtracted photons, the Wigner functions are qualitatively similar, but slight quantitative differences appear, even in the limit of a small reflectivity $\zeta\rightarrow 0$.

Hence, the experimental process produces similar results as the formal definition of the process. Moreover, even for finite $\zeta$, the Wigner functions of the heralded states have negative regions for arbitrary amounts of squeezing.

\begin{figure}[ht]
\centerline{\includegraphics{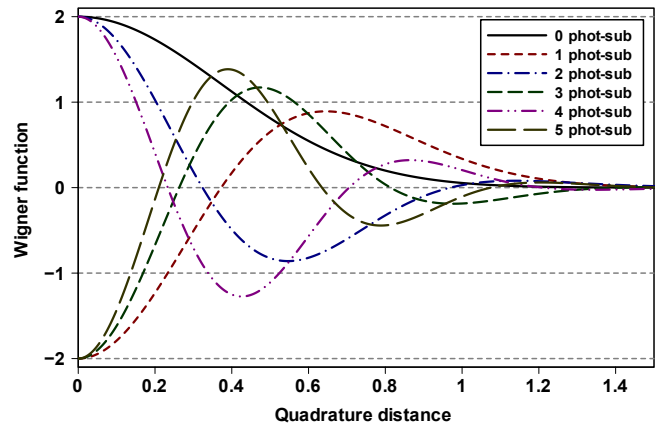}}
\caption{Wigner functions of heralded photon-subtracted squeezed vacuum states ($T=1$) with zero to five subtracted photons, plotted along the squeezed direction, with $\zeta=0.2$. The input squeezed vacuum state has $\avg{n}=1$, so that $A=3$ and $B=B^*=2.83$.}
\label{her1p0}
\end{figure}

\begin{figure}[ht]
\centerline{\includegraphics{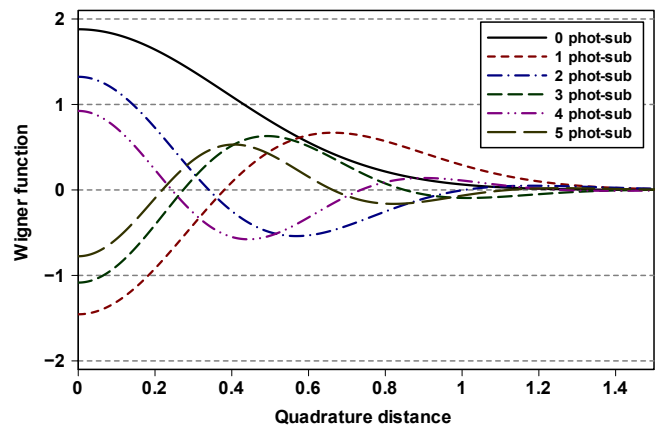}}
\caption{Wigner functions of heralded photon-subtracted squeezed thermal states with zero to five subtracted photons, plotted along the squeezed direction for $\zeta=0.2$, $T=0.9$, $A=3$ and $B=B^*=2.83$.}
\label{her0p9}
\end{figure}

For a quantitative comparison with the formal process, we plot the curves for the heralded photon-subtracted squeezed vacuum states and the heralded photon-subtracted squeezed thermal states with zero to five subtracted photons as functions of $q$ along the line where $p=0$ in figures~\ref{her1p0} and \ref{her0p9}, respectively. The assumed reflectivity of the beam splitter is $\zeta=0.2$. The other parameters are the same as those for the curves in figures~\ref{for1p0} and \ref{for0p9}, respectively. These curves have the same qualitative appearance as those for the formal process, but they are quantitatively different. The effect of a nonzero reflectivity is that it broadens the curves, reducing the effect of the squeezing. By comparing figures~\ref{her0p9} and \ref{for0p9}, we also notice that a nonzero reflectivity slightly reduces the sensitivity with respect to reduced purity in the amplitudes of the oscillations produced by the photon subtractions.

\section{Squeezed Fock states}

Under certain conditions, such as for small squeezing, the Wigner functions of photon-subtracted squeezed vacuum states resemble those of squeezed Fock states. We can use generating functions to investigate such squeezed Fock states.

In this case, the generating function is not associated with a photon subtraction process, but instead with the different Fock states. A generating function for the normalized Wigner functions of the Fock states is given by
\begin{equation}
\mathcal{W}_{{\rm Fock}} = \frac{2}{1+\nu} \exp\left(-2\frac{1-\nu}{1+\nu}|\alpha|^2\right) ,
\label{genfock}
\end{equation}
where $\nu$ is the generating parameter. The Wigner functions of the Fock states are produced by
\begin{equation}
W_{\ket{n}} = \left. \frac{1}{n!} \partial_{\nu}^n \mathcal{W}_{{\rm Fock}} \right|_{\nu=0} .
\label{fockwig}
\end{equation}

The generating function for the Wigner functions of squeezing Fock states is obtained by applying a Bogoliubov transformation, as discussed in section~\ref{bogo}, to the generating function for the Wigner functions of the Fock states in equation~(\ref{genfock}). The result reads
\begin{equation}
\mathcal{W}_{{\rm sq-Fock}} = \frac{2}{1+\nu}\exp\left[-\frac{1-\nu}{1+\nu}\left(2|\alpha|^2A+\alpha^2 B^*+\alpha^{*2} B\right)\right] ,
\label{gensqfock}
\end{equation}
where $A$ and $B$ are defined in equation~(\ref{defab}).

The Wigner function of the squeezed single-photon Fock state is given by the same expression as the single-photon-subtracted squeezed vacuum state in equation~(\ref{eenfs}) with $T=1$. For larger numbers of photons, the expressions are different.

\begin{figure}[ht]
\centerline{\includegraphics{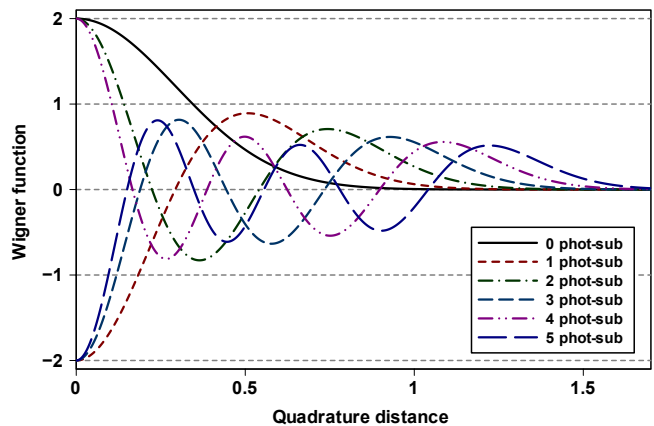}}
\caption{Wigner functions of squeezed Fock states with the same squeezing as in figure~\ref{for1p0}, for zero to five initial photons (prior to squeezing), plotted along $q$ for $p=0$.}
\label{fock}
\end{figure}

In figure~\ref{fock}, we show the curves for the Wigner functions of squeezed Fock states along the line where $p=0$ for the first six Fock states, including the vacuum state. The amount of squeezing is such that the squeezed vacuum state has $\avg{n}=1$. The curves in figure~\ref{fock} show that the Wigner functions of these squeezed Fock state produce qualitatively similar positive and negative regions as found for the photon-subtracted squeezed vacuum states shown in figure~\ref{for1p0}, but apart from the vacuum and single photon Fock states, they are quantitatively different.


\section{\label{fotsubstat}Photon statistics}

\subsection{Generating function}

The properties of a quantum state can be revealed by measuring the photon statistics of the state. The photon statistics is given by the probability distribution of the photon number in the state. The projection operators, their generating function and its Wigner function are provided in section~\ref{projgen}.

To obtain a generating function for the photon statistics of a state, we compute the trace of the state, multiplied by the generating function for the photon number projection operators. In terms of Wigner functions, the generating function for the photon statistics of the photon-subtracted states is obtained by multiplying the generating function for the photon-subtracted states by the generating function for the projection operators, given in equation~(\ref{wigproj}), but replacing $J\rightarrow K$ to have different generating parameters for the photon subtractions and the photon statistics, respectively. Then we integrate over $\alpha$ to obtain a double generating function for both the number of subtracted photons and the photon statistics of such a state.

\subsection{Formal photon-subtracted squeezed thermal states}

In the case of the formal photon subtraction process applied to the squeezed thermal state, the generating function for the photon statistics is obtain by multiplying equation~(\ref{gensqth}) by equation~(\ref{wigproj}) and integrate over $\alpha$. The result is
\begin{eqnarray}
\fl \mathcal{R}_{{\rm fs}} = & \frac{2T}{\sqrt{(1+K)^2T^2+2(1-K^2)TA+(1-K)^2}} \nonumber \\
\fl & \times \exp\left[-\frac{-2TK|\eta|^2 A+T\eta^2 B^*+T\eta^{*2} B-(1-K-T^2-T^2K)|\eta|^2}{(1+K)^2T^2+2(1-K^2)TA+(1-K)^2}\right] .
\label{forthstat}
\end{eqnarray}
It still needs to be normalized with the aid of equation~(\ref{trsqth}).

The expression in equation~(\ref{forthstat}) is a double generating function for the photon statistics of a squeezed thermal state with an arbitrary number of subtracted photons. The number of subtracted photons is determined by the number with derivatives with respect to $\eta$ and $\eta^*$. The probability for observing a given number of photons in the resulting state is determined by the number with derivatives with respect to $K$.

This double generating function for the photon statistics can be used to obtain a generating function for the average number of photons in a photon-subtracted squeezed thermal state. It is given by
\begin{eqnarray}
\mathcal{M}_{{\rm fs}} & \equiv \left. \partial_K \mathcal{R}_{{\rm fs}} \right|_{K=1} \nonumber \\
& = \left[\frac{A-T}{2T}+\frac{A^2-1+(A-T)^2}{4T^2}|\eta|^2 -\frac{A-T}{4T^2}\eta^2 B^*-\frac{A-T}{4T^2}\eta^{*2} B\right] \nonumber \\
& ~~~ \times \exp\left(\frac{2|\eta|^2 (A-T)-\eta^2 B^*-\eta^{*2} B}{4T}\right) .
\end{eqnarray}

For the case with no subtracted photons, the average number of photons is
\begin{equation}
\avg{n} = \frac{A-T}{2T} .
\label{avgn}
\end{equation}
For $A=1$, we obtain the relationship between the purity of a thermal state and its average number of photons, and for $T=1$ we obtain a similar relationship for the squeezed vacuum state.

\subsection{Heralded photon-subtracted squeezed thermal states}

Applying the calculation of the generating function for the photon statistics to the generating function for the heralded photon-subtracted squeezed thermal states, we obtain a double generating function given by
\begin{eqnarray}
\mathcal{R}_{{\rm hs}} & = 2T\left\{\left[1-(K-\zeta K+\zeta J)^2\right]\left(2TA-1-T^2\right) \right. \nonumber \\
& ~~~~ \left. -2(J-K)\left(1-T^2\right)\zeta +2\left(1-K+T^2+T^2K\right)\right\}^{-1/2} .
\label{herthstat}
\end{eqnarray}

The generating function for the average number of photons in a heralded photon-subtracted squeezed thermal state is
\begin{eqnarray}
\fl \mathcal{M}_{{\rm hs}} & \equiv \left. \partial_K \mathcal{R}_{{\rm hs}} \right|_{K=1} \nonumber \\
\fl & = \frac{2T(1-\zeta)\left[(1-\zeta+\zeta J)^2\left(2TA-1-T^2\right)+1-T^2\right]}
 {\left\{\left[1-(1-\zeta+\zeta J)^2\right]\left(2TA-1-T^2\right)+2(1-J)\left(1-T^2\right)\zeta+4T^2\right\}^{3/2}} .
\end{eqnarray}
In the limit of small reflectivity, the average number of photons for no subtracted photons is the same as given in equation~(\ref{avgn}).

\section{Conclusions}

Once the development of quantum information technology moves beyond the proof-of-principle stage and the detailed design of quantum information systems becomes more important, the analyses of such systems will need to address all aspects of physical implementations. Here, we provide analytical tools to aid such detailed quantitative analyses. Although, we consider only the particle-number degrees of freedom here, generating functions can also be used in analyses that incorporate the spatiotemporal degrees of freedom. Such analyses are generally much more complex and the use of generating functions should alleviate the complexity of the calculation in such comprehensive analyses.

The use of generating functions makes it possible to compute the Wigner functions for an arbitrary number of photon subtractions. It allows different models to be investigated, including the formal photon subtractions, squeezed Fock states and the experimentally implemented heralding process to produce such states through photon detection with or without photon-number-resolving detectors.

The generating functions also allow one to compute the photon statistics of such states as generating functions of generating functions. They facilitate the direct calculations of the average number of photons in the states or the variance and standard deviation in the photon number.

Generating functions for the marginal probability distributions can be directly obtained via the integration of the generating functions for the Wigner functions. The complex parameter $\alpha$ is expressed in terms of $q$ and $p$, as defined in equation~(\ref{anaqp}). The resulting expression of the generating function is then integrated over either $q$ or $p$ to obtain a generating function for the marginal probability distributions as a function of the remaining quadrature variable.

\section*{Acknowledgement}

This work was supported in part by funding from the National Research Foundation of South Africa (Grant Numbers: 118532).


\section*{References}


\end{document}